\documentclass[aps,amssymb,showpacs,twocolumn]{revtex4}
\usepackage{graphicx}
\usepackage{color}

\begin{document}

\title{ Meson Scattering in a Pion Superfluid }
\author{\normalsize{Shijun Mao and Pengfei Zhuang }}
\affiliation{Physics Department, Tsinghua University, Beijing
100084, China}
\date{\today}

\begin{abstract}
Instead of the fermion-fermion scattering which identifies the
BCS-BEC crossover in cold atom systems, boson-boson scattering is
measurable and characterizes the BCS-BEC crossover at quark level.
We study $\pi$-$\pi$ scattering in a pion superfluid described by
the Nambu--Jona-Lasinio model. We found that the scattering
amplitude drops down monotonically with decreasing isospin density
and finally vanishes at the boundary of the phase transition. This
indicates a BCS-BEC crossover in the pion superfluid.
\end{abstract}

 \maketitle

There are two kinds of condensed states in usual fermion gas, the
Bardeen--Cooper--Shrieffer condensation (BCS) of fermions where the
pair size is large and the pairs overlap each other, and the
Bose--Einstein condensation (BEC) of molecules where the pair size
is small and the pairs are distinguishable. The BCS wave function
can be generalized to arbitrary attraction which leads to a smooth
crossover from BCS to BEC~\cite{eagles,leggett}. In cold atom systems, the
experimental observable to identify the BCS-BEC crossover is the
$s$-wave scattering between two fermions~\cite{courteille,greiner,zwieriein,bourdel}.

Recently the study on quantum chromodynamics (QCD) phase structure
is extended to finite isospin density. For a QCD system at finite
temperature and baryon and isospin density, the phase transitions
include not only color deconfinement~\cite{hwa}, chiral symmetry
restoration~\cite{hwa} and color superconductor~\cite{alford,rapp}, but also
pion superfluid~\cite{son,he1}. The increasing isospin density
induces a phase transition from normal nuclear matter to pion
superfluid, due to the spontaneous isospin symmetry breaking. By
analogy with the usual superfluid, the BCS-BEC crossover in pion
superfluid can be theoretically described~\cite{matsuo,margueron,mao,sun,mu1} by the
quark chemical potential which is positive in BCS and negative in
BEC, the size of the Cooper pair which is large in BCS and small in
BEC, and the scaled pion condensate which is small in BCS and large
in BEC. However, unlike the fermion-fermion scattering in cold atom
systems, quarks are unobservable degrees of freedom, and thus the
quark-quark scattering can not be measured or used to experimentally
identify the BCS-BEC crossover.

In pion superfluid, the pairs themselves, namely the pion mesons,
are observable objects. One can measure the $\pi-\pi$ scattering to
probe the properties of the pion condensate and in turn the BCS-BEC
crossover. Since pions are Goldstone modes corresponding to the
chiral symmetry spontaneous breaking, the $\pi-\pi$ scattering
provides a direct way to link chiral theories and experimental data
and has been widely studied in many chiral
models~\cite{gasser,bijnens,schulze,quack,huang}. Note that pions are also the
Goldstone modes of the isospin symmetry spontaneous breaking, the
$\pi-\pi$ scattering should be a sensitive signature of the pion
superfluid phase transition.

While the perturbative QCD can well describe the properties of the
new phases at extremely high temperature and density, the study on
the phase transitions at moderate temperature and density depends on
lattice QCD calculations~\cite{kogut} and effective models with QCD
symmetries. One of the widely used effective models is the
Nambu--Jona-Lasinio (NJL) model~\cite{nambu}, which is originally
inspired by the BCS theory and its version at quark
level~\cite{vogl,klevansky,volkov,hatsuda,buballa} gives simple and direct description of the
dynamic mechanisms of spontaneous chiral symmetry breaking, color
symmetry breaking and isospin symmetry breaking. The $s$-wave
$\pi-\pi$ scattering calculated~\cite{schulze,quack,huang} in the model is
consistent with the Weinberg limit~\cite{weinberg} and the
experimental data~\cite{pocanic} in vacuum. In this work, we extend the
calculation to finite isospin chemical potential and focus on its
relation to the BCS-BEC crossover in the pion superfluid.

The Lagrangian density of the two flavor NJL model at quark level is
defined as~\cite{vogl,klevansky,volkov,hatsuda,buballa}
\begin{equation}
\label{njl}
{\cal L} =
\bar{\psi}\left(i\gamma^{\mu}\partial_{\mu}-m_0+\gamma_0 \mu
\right)\psi +G\Big[\left(\bar{\psi}\psi\right)^2+\left(\bar\psi
i\gamma_5\tau_i\psi\right)^2\Big]
\end{equation}
with scalar and pseudoscalar interactions corresponding to $\sigma$
and $\pi$ excitations, where $m_0$ is the current quark mass, $G$ is
the four-quark coupling constant with dimension GeV$^{-2}$, $\tau_i\
(i=1,2,3)$ are the Pauli matrices in flavor space, and $\mu
=diag\left(\mu_u,\mu_d\right)=diag\left(\mu_B/3+\mu_I/2,\mu_B/3-\mu_I/2\right)$
is the quark chemical potential matrix with $\mu_u$ and $\mu_d$
being the $u$- and $d$-quark chemical potentials and $\mu_B$ and
$\mu_I$ the baryon and isospin chemical potentials. At $\mu_I=0$,
the Lagrangian density has the symmetry of $U_B(1)\bigotimes
SU_I(2)\bigotimes SU_A(2)$, corresponding to baryon, isospin and
chiral symmetry. At $\mu_I\neq 0$, the symmetries $SU_I(2)$ and
$SU_A(2)$ are firstly explicitly broken down to $U_I(1)$ and
$U_A(1)$, and then the nonzero pion condensate leads to a
spontaneous breaking of $U_I(1)$, with pions as the corresponding
Goldstone modes. At $\mu_B=0$, the Fermi surface of $u (d)$ and
anti-$d(u)$ quarks coincide and hence the condensate of $u$ and
anti-$d$ is favored at $\mu_I>0$ and the condensate of $d$ and
anti-$u$ quarks is favored at $\mu_I<0$. Finite $\mu_B$ provides a
mismatch between the two Fermi surfaces and will reduce the pion
condensation.

Introducing the chiral and pion condensates
$\sigma=\langle\bar\psi\psi\rangle$ and $\pi=\langle\bar\psi
i\gamma_5\tau_1\psi\rangle$ and taking them to be real, the quark
propagator ${\cal S}$ in mean field approximation can be expressed
as a matrix in the flavor space
\begin{equation}
\label{quarkpropagator}
{\cal S}^{-1}(p)= \left(\begin{array}{cc} \gamma^\mu p_\mu+\mu_u\gamma_0-m_q & 2iG\pi\gamma_5\\
2iG\pi\gamma_5 & \gamma^\mu
p_\mu+\mu_d\gamma_0-m_q\end{array}\right)
\end{equation}
with the dynamical quark mass $m_q=m_0-2G\sigma$ generated by the
chiral symmetry breaking. By diagonalizing the propagator, the
thermodynamic potential can be simply expressed as a condensation
part plus a summation of four quasiparticle contributions~\cite{he1}.
The gap equations to determine the condensates $\sigma$ (or quark
mass $m_q$) and $\pi$ can be obtained by the minimum of the
thermodynamic potential.

In the NJL model, the meson modes are regarded as quantum
fluctuations above the mean field. The two quark scattering via a
meson exchange can be effectively expressed at quark level in terms
of quark bubble summation in the random phase approximation
(RPA)~\cite{vogl,klevansky,volkov,hatsuda,buballa}. The quark bubbles are defined as
\begin{equation}
\label{polarization1} \Pi_{mn}(k) = i\int{d^4p\over (2\pi)^4} Tr
\left(\Gamma_m^* {\cal S}(p+k)\Gamma_n {\cal S}(p)\right)
\end{equation}
with indexes $m,n=\sigma,\pi_+,\pi_-,\pi_0$, where the trace $Tr =
Tr_C Tr_F Tr_D$ is taken in color, flavor and Dirac spaces, the four
momentum integration is defined as $\int d^4 p/(2 \pi)^4=i T \sum_j
\int d^3{\bf p}/(2 \pi)^3$ with fermion frequency $p_0=i \omega_j=i
(2j+1)\pi T\ (j=0,\pm 1, \pm 2, \cdots)$ at finite temperature $T$,
and the meson vertices are from the Lagrangian density (\ref{njl}),
\begin{equation}
\label{vertex} \Gamma_m = \left\{\begin{array}{ll}
1 & m=\sigma\\
i\gamma_5 \tau_+ & m=\pi_+ \\
i\gamma_5 \tau_- & m=\pi_- \\
i\gamma_5 \tau_3& m=\pi_0\ ,
\end{array}\right.\ \
\Gamma_m^* = \left\{\begin{array}{ll}
1 & m=\sigma\\
i\gamma_5 \tau_-& m=\pi_+ \\
i\gamma_5 \tau_+ & m=\pi_- \\
i\gamma_5 \tau_3& m=\pi_0\ . \\
\end{array}\right.
\end{equation}
Since the quark propagator ${\cal S}$ contains off-diagonal
elements, we must consider all possible channels in the bubble
summation in RPA. Using matrix notation for the meson polarization
function $\Pi(k)$ in the $4\times 4$ meson space, the meson
propagator can be expressed as
\begin{equation}
\label{mp}
{\cal D}(k)={2G\over 1-2G\Pi(k)}.
\end{equation}

Since the isospin symmetry is spontaneously broken in the pion
superfluid, the original meson modes $\sigma, \pi_+, \pi_-, \pi_0$
with definite isospin quantum number are no longer the eigen modes
of the Hamiltonian of the system, the new eigen modes
$\overline\sigma, \overline\pi_+, \overline\pi_-, \overline\pi_0$
are linear combinations of the old ones, their masses $M_i
(i=\overline\sigma, \overline\pi_+, \overline\pi_-, \overline\pi_0)$
are determined by the poles of the meson propagator at $k_0=M_i$ and
${\bf k=0}$, $\text{det} \left[ 1-2G \Pi(M_i, {\bf
0})\right]=0$~\cite{he1}, and their coupling constants $g_{i
q\overline q}$ are defined as the residues of the propagator at the
poles~\cite{hao}.

The condition for a meson to decay into a $q$ and a $\overline q$ is
that its mass lies above the $q-\overline q$ threshold. From the
pole equation, the heaviest mode in the pion superfluid is
$\overline\sigma$ and its mass is beyond the threshold value.
Therefore, there will be no $\bar\sigma$ mesons at
$\mu_I>\mu^c_I$~\cite{hao}.
\begin{figure}[hbt]
\centering
\includegraphics[width=5cm]{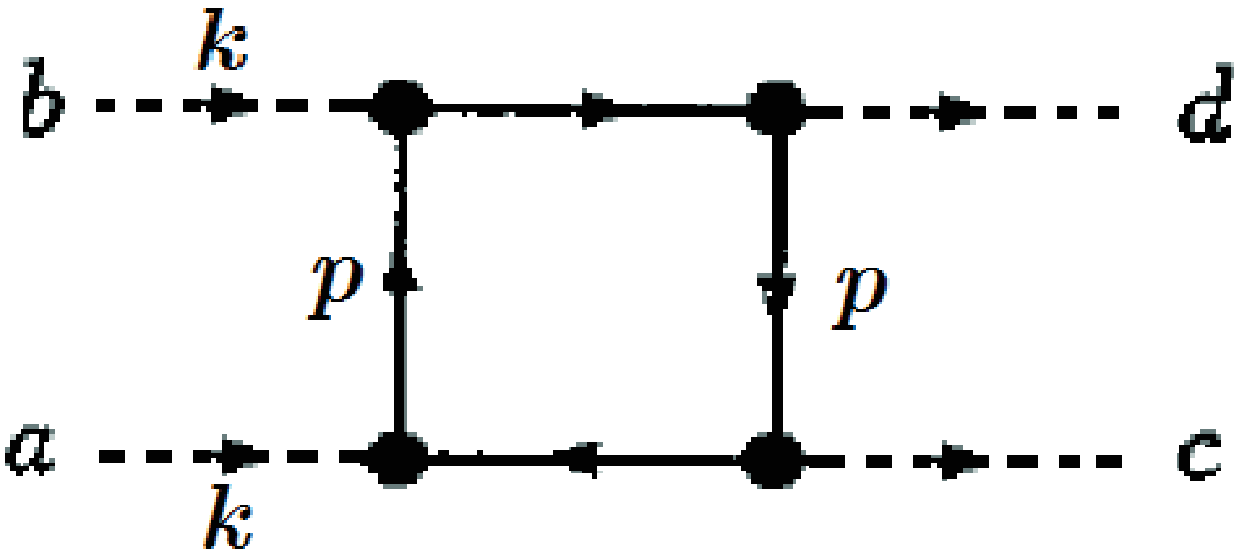}
\includegraphics[width=5.4cm]{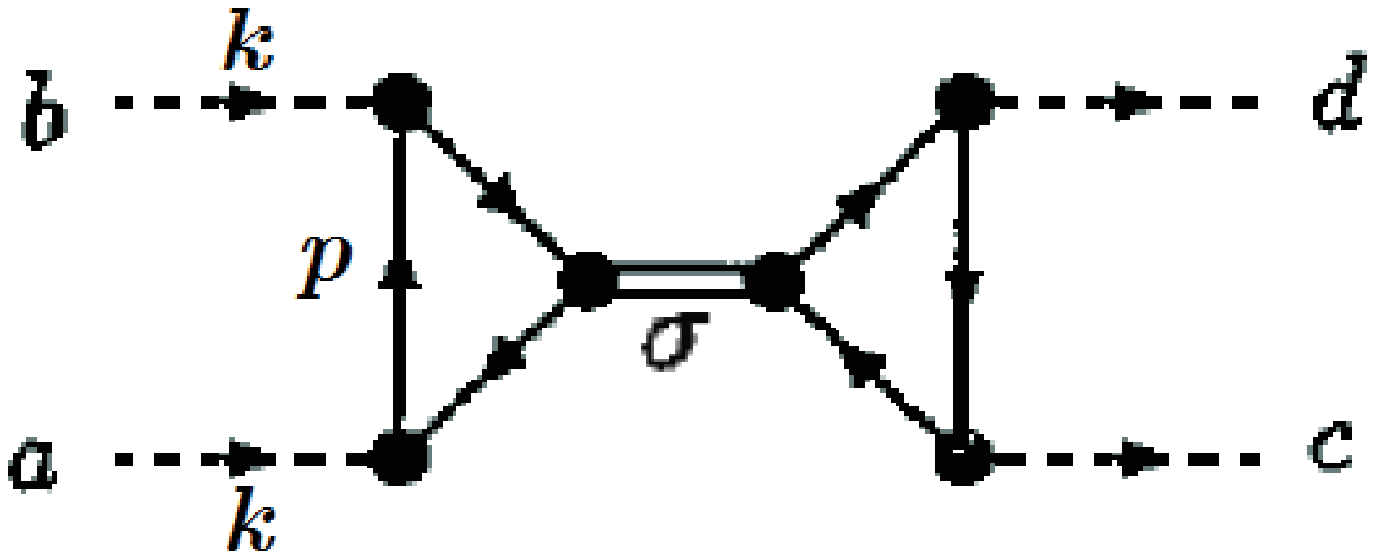}
\caption{The lowest order diagrams for $\pi-\pi$ scattering in the
pion superfluid. The solid and dashed lines are respectively quarks
and pions, and the dots denote meson-quark vertices. } \label{fig1}
\end{figure}

We now study $\pi-\pi$ scattering at finite isospin chemical
potential. To the lowest order in $1/N_c$ expansion, where $N_c$ is
the number of colors, the invariant amplitude ${\cal T}$ is
calculated from the diagrams shown in Fig.\ref{fig1} for the $s$
channel. Different from the calculation in normal
state~\cite{schulze,quack,huang} where both the box and $\sigma$-exchange
diagrams contribute, the $\sigma$-exchange diagrams vanish in the
pion superfluid due to the disappearance of the $\overline\sigma$
meson. This greatly simplifies the calculation in the pion
superfluid.

For the calculation in normal matter at $\mu_I=0$, people are
interested in the $\pi$-$\pi$ scattering amplitude with definite
isospin, ${\cal T}_{I=0,1,2}$, which can be measured in experiments
due to isospin symmetry. However, the nonzero isospin chemical
potential breaks down the isospin symmetry and makes the scattering
amplitude ${\cal T}_{I=0,1,2}$ not well defined. In fact, the new
meson modes in the pion superfluid do not carry definite isospin
quantum numbers. Unlike the chiral dynamics in normal matter, where
the three degenerated pions are all the Goldstone modes
corresponding to the chiral symmetry spontaneous breaking, the pion
mass splitting at finite $\mu_I$ results in only one Goldstone mode
$\overline\pi_+$ in the pion superfluid.

The scattering amplitude for any channel of the box diagrams can be
expressed as
\begin{equation}
\label{tstu} i {\cal T}_{s,t,u}(k) = -2 g_{\overline\pi q \overline
q}^4 \int {d^4 p\over (2 \pi)^4}Tr \prod_{l=1}^4\left[\gamma_5 \tau
{\cal S}_l\right]
\end{equation}
with the quark propagators ${\cal S}_1={\cal S}_3={\cal S}(p)$,
${\cal S}_2={\cal S}(p+k)$, and ${\cal S}_4={\cal S}(p-k)$ for the
$s$ and $t$ channels and ${\cal S}_1={\cal S}_3={\cal S}(p+k)$ and
${\cal S}_2={\cal S}_4={\cal S}(p)$ for the $u$ channel. To simplify
the numerical calculation, we consider in the following the limit of
the scattering at threshold $\sqrt{s}=2M_{\overline\pi}$ and
$t=u=0$, where $s, t$ and $u$ are the Mandelstam variables. In this
limit, the amplitude approaches to the scattering length. Note that
the threshold condition can be fulfilled by a simple choice of the
pion momenta, $k_a=k_b=k_c=k_d=k$ and $k^2=M_{\overline\pi}^2=s/4$,
which facilitates a straightforward computation of the diagrams.
Doing the fermion frequency summation over the internal quark lines,
the scattering amplitude for the process of $\overline \pi_+ \ + \
\overline \pi_+ \rightarrow \overline \pi_+ \ + \ \overline \pi_+$
in the pion superfluid is simplified as
\begin{widetext}
\begin{eqnarray}
\label{totalt} {\cal T}_+ = 18
g_{\overline\pi_+ q \overline q}^4\int {d^3{\bf p}\over (2
\pi)^3}&\Bigg\{&{1\over
E_+^3}\left[\left(f(E_+^-)-f(-E_+^+)\right)-E_+\left(
f'(E_+^-)+f'(-E_+^+)\right)\right]\nonumber\\
&+&{1\over
E_-^3}\left[\left(f(E_-^-)-f(-E_-^+)\right)-E_-\left(f'(E_-^-)+f'(-E_-^+)\right)\right]\Bigg\},
\end{eqnarray}
\end{widetext}
where $E_\pm^\mp=E_\pm \mp\mu_B/3$ are the energies of the four
quasiparticles with
$E_\pm=\sqrt{\left(E\pm\mu_I/2\right)^2+4G^2\pi^2}$ and
$E=\sqrt{{\bf p}^2+m_q^2}$, $f(x)$ is the Fermi-Dirac distribution
function $f(x)=\left(e^{x/T}+1\right)^{-1}$, and $f'(x)=df/dx$ is
the first order derivative of $f$. For the scattering amplitude
outside the pion superfluid, one should consider both the box and
$\sigma$-exchange diagrams. The calculation is straightforward.

Since the NJL model is non-renormalizable, we can employ a hard
three momentum cutoff $\Lambda$ to regularize the gap equations for
quarks and pole equations for mesons. In the following numerical
calculations, we take the current quark mass $m_0=5$ MeV, the
coupling constant $G=4.93$ GeV$^{-2}$ and the cutoff $\Lambda=653$
MeV\cite{zhuang}. This group of parameters correspond to the pion
mass $m_{\pi}=134$ MeV, the pion decay constant $f_{\pi}=93$ MeV and
the effective quark mass $M_q=310$ MeV in the vacuum.
\begin{figure}[hbt]
\centering
\includegraphics[width=7.5cm]{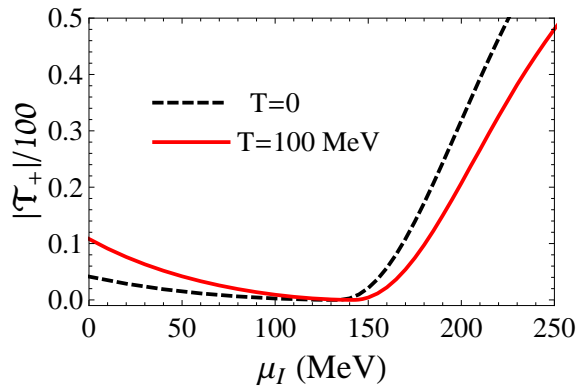}
\caption{ (Color online) The scaled scattering amplitude ${\cal
T}_+$ as a function of isospin chemical potential $\mu_I$ at two
values of temperature $T$. } \label{fig2}
\end{figure}

In Fig.\ref{fig2}, we plot the
scattering amplitude $|{\cal T}_+|$ as a function of isospin
chemical potential $\mu_I$ at two temperatures $T=0$ and $T=100$
MeV, keeping baryon chemical potential $\mu_B=0$. The normal matter with $\mu_I<\mu_I^c$
is dominated by the explicit isospin
symmetry breaking and spontaneous chiral symmetry breaking, and the pion superfluid with $\mu_I>\mu_I^c$ and the
corresponding BEC-BCS crossover is controlled by the spontaneous
isospin symmetry breaking and chiral symmetry restoration. From (\ref{tstu}),
the scattering amplitude is governed by the meson coupling constant,
${\cal T}_+\sim g_{\overline\pi_+ q \overline q}^4$. In the pion
superfluid, the meson mode $\overline\pi_+$ is always a bound state,
its coupling to quarks drops down with decreasing $\mu_I$~\cite{hao},
and therefore the scattering amplitude $\left|{\cal T}_+\right|$
decreases when the system approaches to the phase transition and
reaches zero at the critical value $\mu^c_I$ due to
$g_{\overline\pi_+ q \overline q}=0$ at this point, where the critical isospin chemical potential
$\mu^c_I=m_\pi=134$ MeV at $T=0$ and $142$ MeV at $T=100$
MeV. After crossing the border of the phase
transition, the coupling constant changes its moving trend and start to go up with decreasing isospin chemical potential
in the normal matter~\cite{hao}, and the scattering amplitude smoothly increases and finally approaches its vacuum value for $\mu_I\to 0$.

The above $\mu_I$-dependence of the meson-meson scattering amplitude
in the pion superfluid with $\mu_I>\mu_I^c$ can be understood well
from the point of view of BCS-BEC crossover. We recall that the BCS
and BEC states are defined in the sense of the degree of overlapping
among the pair wave functions. The large pairs in BCS state overlap
each other, and the small pairs in BEC state are individual objects.
Therefore, the cross section between two pairs should be large in
the BCS state and approach zero in the limit of BEC. From our
calculation shown in Fig.\ref{fig2}, the $\pi-\pi$ scattering
amplitude is a characteristic quantity for the BCS-BEC crossover in
pion superfluid. The overlapped quark-antiquark pairs in the BCS
state at higher isospin density have large scattering amplitude,
while in the BEC state at lower isospin density with separable
pairs, the scattering amplitude becomes small. This provides a
sensitive observable for the BCS-BEC crossover at quark level,
analogous to the fermion scattering in cold atom systems.

\begin{figure}[hbt]
\centering
\includegraphics[width=7.5cm]{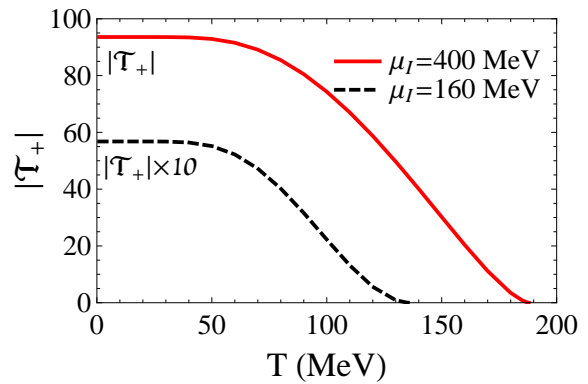}
\caption{(Color online) The scattering amplitude ${\cal T}_+$ as a
function of temperature $T$ at two values of isospin chemical
potential $\mu_I$ in the pion superfluid.} \label{fig3}
\end{figure}
The minimum of the scattering amplitude at the critical point can generally be
understood in terms of the interaction between the two quarks. A strong interaction means a tightly
bound state with small meson size and small meson-meson cross section, and
a weak interaction means a loosely bound state with large meson size
and large meson-meson cross section. Therefore, the minimum of the meson scattering
amplitude at the critical point indicates the most strong quark interaction at the phase transition.
This result is consistent with theoretical calculations for the ratio $\eta/s$~\cite{kovtun,csernai} of
shear viscosity to entropy and for the quark potential~\cite{mu2,jiang}, which show
a strongly interacting quark matter around the phase transition.

With increasing temperature, the pairs will gradually melt and the
coupling constant $g_{\overline\pi q\overline q}$ drops down in the
hot medium, leading to a smaller scattering amplitude at $T=100$ MeV
in the pion superfluid, in comparison with the case at $T=0$, as
shown in Fig.\ref{fig2}. To see the continuous temperature effect on
the scattering amplitude in the BCS and BEC states, we plot in
Fig.\ref{fig3} $\left|{\cal T}_+\right|$ as a function of $T$ at
$\mu_I=160$ and $\mu_I=400$ MeV, still keeping $\mu_B=0$. While the
temperature dependence is similar in both cases, the involved
physics is different. In the BCS state at $\mu_I=400$ MeV,
$\left|{\cal T}_+\right|$ is large and drops down with increasing
temperature and finally vanishes at the critical temperature
$T_c=188$ MeV. Above $T_c$ the system becomes a fermion gas with
weak coupling and without any pair. In the BEC state at $\mu_I=160$
MeV, the scattering amplitude becomes much smaller (multiplied by a
factor of $10$ in Fig.\ref{fig3}). At a lower critical temperature
$T_c=136$ MeV, the condensate melts but the still strong coupling
between quarks makes the system be a gas of free pairs.

In summary, we proposed the meson-meson scattering as a sensitive
probe of the BCS-BEC crossover at quark level. Different from the
fermion-fermion scattering which is often used to experimentally
identify the BCS-BEC crossover in cold atom systems, quark
scattering can not be measured and its function to characterize the
BCS-BEC crossover at quark level is replaced by the meson
scattering. In the BCS quark superfluid, the large and overlapped
pairs lead to a large pair-pair cross section, but the small and
individual pairs in the BEC superfluid interact weakly with small
cross section. In the frame of a two flavor NJL model at finite
temperature and isospin density, we calculated the $\pi-\pi$
scattering amplitude in the pion superfluid. It is large at high
isospin chemical potential and drops down monotonically with
decreasing isospin chemical potential and finally approaches zero at
the border of the pion superfluid, indicating a BCS-BEC crossover.

The meson scattering amplitude $\left|{\cal T}_+\right|$ shown in
Figs.\ref{fig2} and \ref{fig3} are obtained in a particular model,
the NJL model, which has proven to be rather reliable in the study
on chiral, color and isospin condensates at low temperature. Since
there is no confinement in the model, one may ask the question to
what degree the conclusions obtained here can be trusted. From the
general picture for BCS and BEC states, the feature that the meson
scattering amplitude approaches to zero in the process of BCS-BEC
crossover can be geometrically understood in terms of the degree of
overlapping between the two pairs. Therefore, the qualitative
conclusion of taking meson scattering as a probe of BCS-BEC
crossover at quark level may survive any model dependence. Our
result that the molecular scattering amplitude approaches to zero in
the BEC limit is consistent with the recent work for a general
fermion gas~\cite{he2}. Different from a system with finite baryon density
where the fermion sign problem~\cite{muroya} makes it difficult to simulate QCD
on lattice, there is in principle no problem to do lattice QCD calculations
at finite isospin density~\cite{kogut}. From the recent lattice QCD results~\cite{detmold}
at nonzero isospin chemical potential in a canonical
approach, the scattering length in the pion superfluid increases with increasing
isospin density, which qualitatively supports our conclusion here.

\appendix {\bf Acknowledgement:} The work is supported by the NSFC
(Grant Nos. 10975084 and 11079024), RFDP (Grant No.20100002110080
) and MOST (Grant No. 2013CB922000).

\end{document}